\pdfoutput=1
\documentclass[aps,prd,twocolumn,superscriptaddress,preprintnumbers,floatfix,nofootinbib,notitlepage,showkeys,showpacs]{revtex4-1}
\usepackage[utf8]{inputenc}
\usepackage{cancel}

\usepackage{graphicx}
\usepackage{hyperref}
\usepackage{latexsym}
\usepackage{amsmath}
\usepackage{amssymb}
\usepackage{bbm}
\usepackage{microtype}

\usepackage[normalem]{ulem}
\usepackage{pdfsync}
\usepackage{epsfig}
\usepackage{epstopdf}
\usepackage{subfigure}
\usepackage{color}
\usepackage{comment}
\usepackage{slashed}
\usepackage{placeins}
\usepackage{cleveref}

\usepackage{xcolor}
\definecolor{cites}{RGB}{0,180,0}
\definecolor{links}{RGB}{200,0,0}
\hypersetup{colorlinks=true,citecolor=cites,linkcolor=links,urlcolor=blue}

\usepackage{tikz}
\usetikzlibrary{decorations.pathmorphing,decorations.markings}
\usepackage{graphicx}
\tikzset{
photon/.style={decorate, decoration={snake,amplitude=2pt, segment length=5pt}, draw=black},
particle/.style={draw=black, postaction={decorate}, decoration={markings,mark=at position .5 with {\arrow[draw=black]{>}}}},
antiparticle/.style={draw=black, postaction={decorate}, decoration={markings,mark=at position .5 with {\arrow[draw=black]{>}}}},
gluon/.style={decorate, draw=black, decoration={coil,amplitude=4pt, segment length=5pt}}
goldstone/.style={draw=green,postaction={decorate},decoration={markings,mark=at position .5 with {\arrow[draw=blue]{>}}}}
}
\usepackage{tabu}
\usepackage{float}

\newcommand{\intron}[1]{}%{#1}

\newcommand{\be}{\begin{equation}}
\newcommand{\ee}{\end{equation}}
\newcommand{\bea}{\begin{eqnarray}}
\newcommand{\eea}{\end{eqnarray}}
\newcommand{\mh}{m_{h^0}}
\newcommand{\mH}{m_{H^0}}
\begin{document}
\title{Direct and indirect probes of Goldstone dark matter}
\author{Tommi Alanne}
\email{tommi.alanne@mpi-hd.mpg.de}
\affiliation{Max-Planck-Institut f\"{u}r Kernphysik, %\\
    Saupfercheckweg 1, 69117 Heidelberg, Germany}
\author{Matti Heikinheimo}
\email{matti.heikinheimo@helsinki.fi}
\affiliation{Department of Physics, University of Helsinki, %\\
                      P.O.Box 64, FI-00014 University of Helsinki, Finland}
\affiliation{Helsinki Institute of Physics, %\\
                      P.O.Box 64, FI-00014 University of Helsinki, Finland}
\author{Venus Keus}
\email{venus.keus@helsinki.fi}
\affiliation{Department of Physics, University of Helsinki, %\\
                      P.O.Box 64, FI-00014 University of Helsinki, Finland}
\affiliation{Helsinki Institute of Physics, %\\
                      P.O.Box 64, FI-00014 University of Helsinki, Finland}
\author{Niko Koivunen}
\email{niko.koivunen@helsinki.fi}
\affiliation{Department of Physics, University of Helsinki, %\\
                      P.O.Box 64, FI-00014 University of Helsinki, Finland}
\affiliation{Helsinki Institute of Physics, %\\
                      P.O.Box 64, FI-00014 University of Helsinki, Finland}
\author{Kimmo Tuominen}
\email{kimmo.i.tuominen@helsinki.fi}
\affiliation{Department of Physics, University of Helsinki, %\\
                      P.O.Box 64, FI-00014 University of Helsinki, Finland}
\affiliation{Helsinki Institute of Physics, %\\
                      P.O.Box 64, FI-00014 University of Helsinki, Finland}

\begin{abstract}
  There exists a general model framework where dark matter can
be a vanilla WIMP-like thermal relic with a mass of ${\cal O}$(100 GeV),
but it still escapes direct detection. This happens, if the dark matter particle is
a Goldstone boson whose scattering with ordinary matter is suppressed at
low energy due to momentum-dependent interactions. We outline general
features of this type of models and analyse a simple realization of these
dynamics as a concrete example. In particular, we show that although
direct detection of this type of dark matter candidate is very challenging,
the indirect detection can already provide relevant constraints. Future
projections of the indirect-detection experiments allow for even more stringent exclusion limits and can rule out models of this type.
\end{abstract}

%\keywords{}

\preprint{HIP-2018-34/TH}

\maketitle

%%%%%%%%%%%%%%%%%%%%%%%%%%%%%%%%%%%%%%%%%%%%%%%%%%%%%%%%%%%%%%%%%%%%%%%%%%%%%%%%%%%%%%%%%%%%%%%%%%%%
\section{Introduction}
%%%%%%%%%%%%%%%%%%%%%%%%%%%%%%%%%%%%%%%%%%%%%%%%%%%%%%%%%%%%%%%%%%%%%%%%%%%%%%%%%%%%%%%%%%%%%%%%%%%%
Over the recent years the paradigm of dark matter (DM) as a thermal relic has
been heavily challenged by the results from direct-detection
experiments, which systematically provide more stringent bounds on the strength of the interaction between dark and ordinary matter~\cite{Akerib:2017kat, Cui:2017nnn, Aprile:2018dbl}. One possible interpretation of these results is
that DM constitutes a secluded sector which is very feebly coupled with the
Standard Model (SM). In this type of freeze-in models~\cite{McDonald:2001vt,Hall:2009bx,Bernal:2017kxu} the portal coupling of ${\cal O}(10^{-10})$ allows the observed DM
abundance to be produced out of equilibrium without ever equilibrating
with the SM heat bath. Consequently, models of this type are usually
completely invisible in the direct-detection experiments.

However, several solutions to address this issue also within the standard thermal-relic paradigm have been proposed.
For example, a large enough particle
content beyond the SM could make the cross sections probed in direct-detection experiments
distinct from the cross sections affecting the freeze-out dynamics in the
early universe~\cite{Pospelov:2007mp}.
Such non-minimal hidden-sector models, on the other hand, generally extend the
usual problems of explaining the origins and hierarchies of the required
mass scales and couplings.

It would therefore be worthwhile to outline general features of
models where similar effects could be achieved with minimal particle
content. In this paper, we investigate simple scalar extensions of the
SM where the DM arises as a pseudo-Goldstone boson of an approximate
global symmetry; examples of such models have been studied recently in
e.g.~\cite{Alanne:2014kea,Gross:2017dan,Azevedo:2018oxv, Azevedo:2018exj,Ishiwata:2018sdi,
Balkin:2018tma,Alanne:2018xli}. Since generally the
interactions of Goldstone bosons are momentum suppressed, this will naturally
suppress the DM-nucleon elastic scattering cross section relevant for direct search experiments, which
operate in the limit of zero momentum transfer. On the other hand, the annihilation cross section of pseudo-Goldstone bosons does not vanish in the non-relativistic limit, as the momentum transfer in this process contains a non-zero contribution from the rest-mass of the annihilating particles.
Therefore, in this class of models, DM with the observed abundance can be produced as a thermal relic, while the direct-detection cross section is small enough for the DM to have escaped all present direct searches. 

The indirect DM searches, on the other hand, probe the same annihilation process that is relevant for determining the DM abundance. Since this scattering amplitude does not vanish in the limit of zero incoming three-momentum, the indirect-detection signal expected from a pseudo-Goldstone DM particle is of a similar magnitude as that of a generic WIMP. Indeed, indirect-detection limits turn out to be very constraining for such DM candidates, and future prospects for observing pseudo-Goldstone DM up to couple hundred GeV masses in indirect searches are promising.

There is, however, a caveat to the above simple picture: to account for a
finite mass of the
Goldstone boson DM candidate, an explicit breaking of the global
symmetry is required. Since the vanishing of the direct-detection cross section at zero momentum transfer rests on the foundation of the underlying global symmetry and the resulting Goldstone nature of the DM particle, it should not be expected to strictly hold in the presence of explicit symmetry breaking. This issue has been investigated in the recent works~\cite{Azevedo:2018exj,Ishiwata:2018sdi,Balkin:2018tma}, where one-loop contributions to the direct-detection amplitude were considered. It turns out indeed that the one-loop contribution does not vanish in the zero-momentum transfer limit, and the resulting prediction for the direct-detection event rate is considerably larger than the tree-level prediction. However, for most parts of the parameter space, the pseudo-Goldstone DM still remains concealed below the neutrino floor
and thus hidden from direct detection.

To demonstrate these mechanisms and effects quantitatively, we will use an
effective model with O($N$)/O($N-1$) symmetry-breaking pattern as an example.
However, our analysis can be applied to all models where the hidden sector degrees of freedom consist 
of Goldstone bosons and one massive scalar mixing with the Higgs boson.
We will discuss how the effective theory with the leading symmetry breaking
operators of dimension two
and four may emerge from simple high-energy dynamics, thus approaching the
issue of accounting for the effects of explicit symmetry breaking from a
top-down direction, and compare the results to the bottom-up approach of
computing the radiative corrections in the low-energy effective
theory~\cite{Gross:2017dan,Azevedo:2018exj,Ishiwata:2018sdi,Balkin:2018tma}.
We will
present a compilation of current experimental constraints from direct and
indirect observations on DM scattering as well as from collider
experiments.

The paper is organized as follows: In Sec.~\ref{sec:constraints} we will outline the general model framework and relevant constraints.
Then, in Sec.~\ref{sec:model} we will carry out a detailed quantitative
analysis for single component Goldstone DM and show how various experimental and observational constraints operate.
In Sec.~\ref{sec:checkout} we will present our conclusions and outlook for further work. Some analytic results relevant for the analysis and the non-linear representation of the model are collected
in appendices.

%%%%%%%%%%%%%%%%%%%%%%%%%%%%%%%%%%%%%%%%%%%%%%%%%%%%%%%%%%%%%%%%%%%%%%%%%%%%%%%%%%%%%%%%%%%%%%%%%%%%
\section{General model framework and constraints}
\label{sec:constraints}

We consider a scenario where DM arises as a pseudo-Goldstone boson associated
with breaking of an approximate global symmetry. Such a DM
candidate emerges naturally, for example, in models featuring new strong dynamics
leading to a spectrum of mesons at low energies (see e.g. Refs~\cite{Hur:2007uz,Hur:2011sv,Bhattacharya:2013kma,Heikinheimo:2013fta}). To keep the discussion simple, and to establish generic features of this
type of DM, we will here
treat the effective theory degrees of freedom as elementary fields.

Concretely, we consider extending the SM with a scalar field, $S$, transforming under some
irreducible $N$-dimensional representation of a global symmetry $G$.
The scalar field is assumed to develop a vacuum expectation value (vev), $\langle S\rangle=w$, such that the symmetry breaks to a subgroup $G'$.
We consider explicitly a class of models with the symmetry
breaking pattern O($N$)/O($N-1$). Then the relevant low energy degrees
of freedom are parametrized in terms of an O($N$) vector
\be
\Sigma=(\eta_1,\eta_2,\dots,\eta_{N-1},\sigma),
\ee
where $\sigma$ is the field direction
along which the vev of $\Sigma$ develops.
Then, the effective Lagrangian for the singlet sector and
its interactions with the visible sector is
\be
{\cal L}= \frac{1}{2} \partial_\mu\sigma\partial^\mu\sigma+
\frac{1}{2}\partial\eta_a\partial\eta_a-V(\Sigma,H),
\label{eq:generaltheory}
\ee
where $H$ is the usual SM Higgs field.
The potential $V(\Sigma,H)$ is given by
\begin{align}
\label{eq:scalarpot}
V(\Sigma,H)=&\,\mu_H^2H^\dagger H+\frac{1}{2}\mu_\Sigma^2\Sigma^\dagger\Sigma
+\lambda_H(H^\dagger H)^2\nonumber\\
&+\frac{\lambda_{H\Sigma}}{2}(H^\dagger H)\Sigma^\dagger\Sigma
+\frac{\lambda_\Sigma}{4} (\Sigma^\dagger\Sigma)^2
+V_{\rm{sb}},
\end{align}
where $V_{\mathrm{sb}}$ contains the symmetry-breaking contributions.

The field $\eta^a$ will be a DM candidate as it is protected against decay by 
a symmetry within $G'$. Of course, to constitute cold DM of thermal origin, it must be massive and the symmetry $G$ must be explicitly broken. On the effective Lagrangian level, one could include the leading symmetry breaking term $M_\eta^2 \eta_a\eta_a$,
which would arise from some higher scale physics coupling with $\Sigma$ and
integrated out to obtain the low energy effective theory, Eq.~\eqref{eq:generaltheory}.
A simple possibility would be a heavy fermion $\Psi$ coupling with all components $\eta^a$ via
\be
\Delta{\cal L}_{\rm{UV}} \supset ig_P\eta^a\bar{\Psi}\gamma_5\Psi.
\ee
When integrated out, this heavy fermion yields a non-zero mass for all Goldstone bosons, $M_\eta^2\sim g_P^2 M_\psi^2$. Furthermore,
contributions to four point couplings between components $\Sigma^a$ are generated, but these are parametrically smaller, of order
$\sim g_P^4$.

Taking into account the coupling with the Higgs would lead to further
explicit symmetry breaking terms of the form
$\lambda_X H^\dagger H(\sigma^2-\eta_a\eta_a)$, where $\lambda_X\sim \lambda_{H\Sigma}g_P^4$. We will therefore take
\be
V_{\rm{sb}}=-\frac{1}{2}\mu_X^2(\sigma^2-\eta_a\eta_a)-\frac{\lambda_X}{2} (H^\dagger H)(\sigma^2-\eta_a\eta_a).
\label{eq:sbreaking}
\ee

With these definitions, we can now see how
the essential dynamics for DM scattering on ordinary matter can be inferred from the Lagrangian, Eq.~\eqref{eq:generaltheory}. As $\Sigma$ and $H$ obtain
their vevs, $\langle \Sigma\rangle=w$ and $\langle H\rangle=v$, the effective potential in Eq.~\eqref{eq:scalarpot} is minimised by requiring
\begin{align}
    \mu_H^2 &=-\frac{1}{2}(2\lambda_H v^2+(\lambda_{H\Sigma}-\lambda_X)w^2),\nonumber \\
    \mu_\Sigma^2 &=-\frac{1}{2}(2\lambda_\Sigma w^2+(\lambda_{H\Sigma}-\lambda_X)v^2-2\mu_X^2).
    \label{eq:minconds}
\end{align}
The scalar field $\sigma$ will mix with the neutral component the doublet $H$; the mass eigenstates are the physical Higgs boson, $h^0$,
and another massive scalar, $H^0$.
These mass eigenstates are given by
\begin{equation}
    \label{eq:rotation}
    \left(\begin{array}{c}h^0\\H^0\end{array}\right)=
    \left(\begin{array}{cc}\cos\alpha & -\sin\alpha\\ \sin\alpha & \cos\alpha\end{array}\right)
    \left(\begin{array}{c}\phi\\ \sigma\end{array}\right),
\end{equation}
with
\begin{equation}
    \label{eq:t2a}
    \tan(2\alpha)=-\frac{(\lambda_{H\Sigma}-\lambda_X)vw}{\lambda_H v^2-\lambda_\Sigma w^2}.
\end{equation}
Now we can solve $\lambda_\Sigma$ from Eq.~\eqref{eq:t2a}, and trade $\lambda_H$, $w$, and $\mu_X^2$ for the
masses of the eigenstates, $\mh,\mH,m_\eta$. Identifying the lightest eigenstate with the SM Higgs, ${\mh=125}$~GeV,
leaves us with $\lambda_{H\Sigma}, \mH,m_\eta$, $\alpha$, and $\lambda_X$ as input parameters.

The interactions between SM matter and DM arise from the SM Yukawa couplings providing the usual link between the Higgs doublet and ordinary matter, and from the coupling of the DM with both the Higgs and the singlet scalar, $\sigma$. Due to the mixing of the scalar, the scattering of DM
on ordinary matter is mediated by both scalar mass eigenstates, $h^0$ and $H^0$.

Generally, the scattering of Goldstone bosons is proportional to the momentum transfer in the process, and one expects that the DM scattering cross section on ordinary matter becomes suppressed. In the model setup described above,
neglecting $\lambda_X$, which we have argued to be suppressed,
one finds that
\begin{align}
\frac{\mathrm{d}\sigma_{\mathrm{SI}}}{\mathrm{d}\cos\theta} \sim \frac{\lambda_{HS}^2f_N^2m_N^2}{(\mh^2-t)^2(\mH^2-t)^2} t^2.
\end{align}
The direct-detection cross section vanishes as $t\rightarrow 0$.

However, there is an important refinement of this argument:
the effects of the symmetry breaking contributions we have introduced above,
are ${\cal O}(t^0)$, rather
than vanishing proportionally to the momentum transfer.
As a result, the
symmetry breaking contributions can quantitatively be significant, even if generated at one-loop or higher order, and must
be treated with care. We study this in detail 
in Sec.~\ref{sec:model}, where we will also quantify the effects
arising from the non-zero values of the coupling $\lambda_X$.

Let us then consider the general features of the constraints from collider searches and cosmological and astrophysical observations, that pertain to the pseudo-Goldstone DM model. First, the observed abundance of DM must be produced as a thermal relic. This implies that the thermal annihilation cross section must satisfy
$\langle \sigma v_{\mathrm{rel}}\rangle_0\simeq 3\cdot 10^{-26}\,\mathrm{cm}^3/s$.
We focus on a scenario where the DM is the lightest non-SM particle, and consider the annihilation cross section to SM final states. 

Second, models with a hidden sector coupling to the visible sector via a scalar portal are constrained by collider experiments in two respects: 
The mass eigenstates are mixtures of the neutral component of the SU(2) scalar doublet and the singlet scalar.
The associated mixing angle, $\alpha$,
is constrained from the Higgs couplings measurements by $\sin\alpha\lesssim 0.3$~\cite{Ilnicka:2018def}.
Additionally, if the hidden sector contains states which are
lighter than half of the mass of the Higgs boson, they will contribute to
the invisible Higgs decays. Such decays are currently bounded by $\mathrm{Br}(h^0\to \mathrm{inv})\leq 0.23$~\cite{Aad:2015pla,Khachatryan:2016whc}.

Third, direct-detection experiments provide stringent exclusion bounds for a vanilla scalar DM with mass of ${\cal{O}}$(100 GeV). However, in the framework studied here, the Goldstone nature of the DM particle relaxes the direct-detection bounds due to the momentum suppression of the cross section, as discussed above. 
Generally, in this scenario the WIMP couples to the nucleus via the Higgs boson, $h^0$, and the heavier mass eigenstate, $H^0$. The strength of the interaction depends on the mixing pattern of the scalars and whether the WIMP is a scalar or a fermion.
In both cases, the Higgs-nucleon coupling is of the form $f_N m_N /v$, with $m_N=0.946$ GeV, where we neglect the small differences between neutrons and protons.
The effective Higgs-nucleon coupling,
\begin{equation}
f_N\equiv \frac{1}{m_N}\sum_q\langle N|m_q\bar{q}q| N\rangle,
\end{equation}
describes the normalized total quark-scalar current within the nucleon. The quark currents of the nucleon have been the subject of intensive lattice studies, supplemented by chiral perturbation theory methods and pion nucleon scattering analysis. Consequently, the current value for $f_N\simeq0.3$~\cite{Alarcon:2011zs,Alarcon:2012nr,Cline:2013gha} is fairly well determined.
The spin-independent cross section for a WIMP scattering on nuclei is computed by considering the $t$-channel exchange of $h^0$ and $H^0$. Due to the Goldstone nature of the DM candidate, even in the presence of the symmetry breaking operators, this cross section is suppressed, allowing for the compatibility with the current direct search limits. We will show this in detail for an explicit model in the next section.

Finally, the model is constrained by indirect-detection. The indirect-detection experiments attempt to observe the annihilation products of DM particles originating from regions of high DM number density in the cosmos, such as the central regions of DM halos. The non-observation of such signals leads to an upper limit on the DM annihilation cross section. Currently the most constraining limits for our mass range of interest are from Fermi-LAT observations of dwarf spheroidal satellite galaxies of the Milky Way~\cite{Fermi-LAT:2016uux}.
While the direct-detection cross section is strongly suppressed at low momentum transfer for Goldstone DM, this is not generally true for the annihilation cross section. Indeed, in the simplest Goldstone-DM models, the annihilation amplitude is an $s$ wave, so that a non-zero amplitude at zero incoming three-momentum exists, making indirect detection a promising avenue for observing Goldstone DM.

We will quantify all the above constraints in the next section for an explicit
model of Goldstone DM.

\section{A model example}
\label{sec:model}
%%%%%%%%%%%%%%%%%%%%%%%%%%%%%%%%%%%%%%%%%%%%%%%%%%%%%%%%%%%%%%%%%%%%%%%%%%%%%%%%%%%%%%%%%%%%%%%%%%%%

We will now consider the model introduced in Sec.~\ref{sec:constraints}
in the simplest case of O(2) symmetry. 
This scenario is equivalent to the extension of the SM with a complex singlet featuring a global U(1) symmetry~\cite{Barger:2008jx}.
Then the DM candidate is a single
Goldstone boson. 
Theories where Goldstone DM forms a degenerate multiplet can be easily addressed using our results together with simple scaling: 
the event rate in direct-detection experiments scales linearly with the degeneracy, $g$,  while the rate in indirect detection 
is independent of $g$.

For most of the analysis,
we set symmetry breaking coefficient
$\lambda_X$ in Eq.~\eqref{eq:sbreaking} to zero and focus on the effect of the
leading symmetry breaking term, $\mu_X^2\neq 0$. 
We also show the quantitative effect from $\lambda_X\neq 0$ on our results.

%%RELIC ABUNDANCE
\subsection{Relic abundance}
\label{subsec:relic}
To determine the relic abundance of the scalar $\eta$, we compute the total annihilation cross section into SM particles and
compare it with the standard thermal annihilation cross section, $\langle \sigma v_{\mathrm{rel}}\rangle_0\simeq 3\cdot 10^{-26}\,\mathrm{cm}^3/s$,
as described in Sec.~\ref{sec:constraints}.
The following annihilation channels need to be taken into account: $\eta\eta\rightarrow h^0h^0, VV$ and $\bar{f}f$, where $V$ denotes the electroweak gauge bosons, $V=W,\,Z$, and $f$ are the SM fermions\footnote{We are
interested in the case where $\eta$ is a Goldstone boson and thus lighter than $H^0$. Therefore, we do not consider $\eta\eta\to h^0H^0, H^0H^0$ channels. }.
The annihilation cross sections to these three distinct final states are given in
Appendix~\ref{xsect}.
Following Ref.~\cite{Cline:2012hg}, we include the four-body
final states due to the virtual $W$ and $Z$ exchange using the full width of the Higgs~\cite{deFlorian:2016spz} in the calculation of the annihilation
cross section in the mass range $\mh/2\leq m_{\eta}\leq 100$~GeV.
We show the curve where $\langle \sigma v_{\mathrm{rel}}\rangle=\langle \sigma v_{\mathrm{rel}}\rangle_0$ for $\mH=500,750$\,GeV, $\alpha=0.3$, $\lambda_X=0$ in Figs~\ref{fig:xSecs}
and~\ref{fig:loop} along with the constraints from invisible Higgs decays and direct-detection searches, and in Fig.~\ref{fig:indirect} together with the
indirect-detection bounds.

\subsection{Invisible Higgs decays}
If $m_\eta<\mh/2$,  one needs to take into account the constraints from invisible Higgs decays, currently bound to be
$\mathrm{Br}(h^0\to \mathrm{inv})\leq 0.23$~\cite{Aad:2015pla,Khachatryan:2016whc}. The Higgs total
decay width to the visible SM channels is $\Gamma_{h^0}=4.07$ MeV for $\mh=125$ GeV~\cite{Dittmaier:2011ti}, and
the $h^0\rightarrow \eta\eta$ width is given by
\begin{equation}
\Gamma_{h^0\rightarrow \eta\eta}=\frac{\lambda_{h^0\eta\eta}^2}{32\pi \mh} \sqrt{1-\frac{4m_\eta^2}{\mh^2}},
\label{higgsdecaywidth}
\end{equation}
where
\begin{equation}
    \label{eq:}
    \lambda_{h^0\eta\eta}=\frac{(\lambda_{HS}-\lambda_X)\mh^2v}{\cos\alpha(\mh^2-\mH^2)}+2\lambda_X v \cos\alpha.
\end{equation}
We show the excluded region as a function of DM mass and $\lambda_{HS}$ for $\mH=500, 750$\,GeV, $\alpha=0.3$, $\lambda_X=0$ superimposed with direct-detection
bounds in Figs~\ref{fig:xSecs} and~\ref{fig:loop}. The invisible Higgs decays, exclude the full parameter space for a light DM except for the very narrow resonance region where $m_\eta\approx \mh/2$.

%%DIRECT DETECTION
\subsection{Direct detection}
\paragraph{Tree-level estimate}
At tree level, the spin-independent (SI) direct-detection cross section is
\begin{equation}
\frac{\mathrm{d}\sigma_{\mathrm{SI}}}{\mathrm{d}\cos\theta} = \frac{\lambda_{\mathrm{eff}}^2f_N^2m_N^2\mu_R^2}{8\pi
    m_\eta^2},
    \label{eq:SIlinear}
\end{equation}
where $\mu_R =m_N m_\eta/(m_N+m_\eta)$ is the reduced mass of the $\eta$-nucleon system, and
\begin{align}
  \label{eq:lambdaeff}
    \lambda_{\mathrm{eff}} &=\frac{\lambda_{HS}t}{(\mh^2-t)(\mH^2-t)}\\
    & -\frac{2\lambda_X\left [ \sin^2\alpha (\mh^2-t/2)
	+\cos^2\alpha (\mH^2-t/2)\right ]}{(\mh^2-t)(\mH^2-t)}.\nonumber
\end{align}
Note that the first term vanishes at zero momentum transfer, as expected for a
Goldstone boson. Furthermore, note that this result persists in the presence
of the soft symmetry-breaking mass term for $\eta$; the non-zero contributions
at zero momentum transfer arise only due to the explicit symmetry-breaking
term, $\lambda_X$, in the scalar potential in Eq.~\eqref{eq:scalarpot}.

\begin{figure}[t]
	\includegraphics[width=0.45\textwidth]{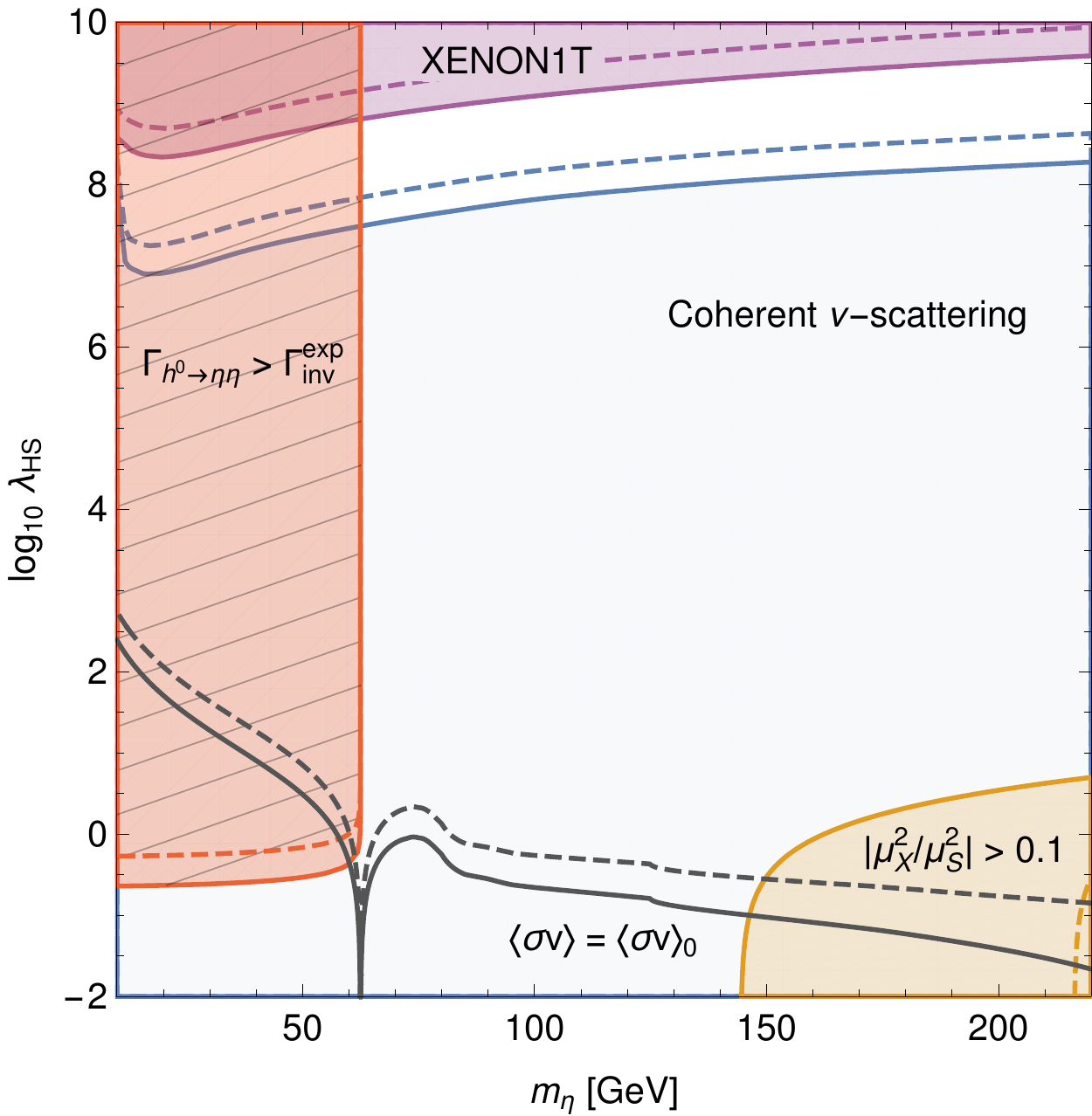}
    \caption{The current limits from XENON1T on the spin-independent scattering
    cross section of DM off of nuclei assuming the
    tree-level momentum-suppressed estimate of the cross sections (upper purple
    shaded region)
    and invisible Higgs decays (red shaded region on left), and the region where the
    SI cross section reaches the coherent neutrino scattering cross section (lower blue region). The limits are
    for $\mH=500$~GeV and $\alpha=0.3$. The dashed curves show the change for $\mH=750$~GeV. The yellow shaded region
    shows where $\mu_X^2/\mu_S^2>0.1$.
    The gray solid (dashed) curve shows where
    the observed DM abundance is achieved for $\mH=500$ GeV ($750$ GeV).
    }
    \label{fig:xSecs}
\end{figure}

We show in Fig.~\ref{fig:xSecs} the current limit from XENON1T experiment~\cite{Aprile:2018dbl}, along with the tentative boundary of the coherent neutrino-scattering cross section~\cite{Billard:2013qya} and the experimental bound for invisible Higgs
decays~\cite{Aad:2015pla,Khachatryan:2016whc} in the $(m_{\eta},\lambda_{HS})$ plane in the case of $\lambda_X =0$
for fixed values of $\mH=500$~GeV and
$\alpha=0.3$. For higher values of $\mH$, the boundaries move upwards in the plane; lowering $\alpha$ has the same
effect on the invisible-Higgs-decay boundary. The dashed curves show the
case of $\mH=750$~GeV.
We have chosen a constant incoming velocity of the DM particles\footnote{We have checked for a sample of parameter points that integrating over the DM velocity distribution given by the standard halo model gives similar results.}, $v_\eta=10^{-3} c$.  The gray solid (dashed) curve shows where the
observed DM abundance is obtained for $\mH=500$ GeV ($750$ GeV); the whole curve lies beneath the neutrino floor.
Note that the highly
non-perturbative values of $\lambda_{HS}$ in the figure are shown for illustration only. The allowed region of the perturbatively computed relic density curve lies in
the domain where the use of perturbation theory is justified. The yellow shaded region in the lower right corner shows where the
explicit-breaking mass term, $\mu_X^2$, starts to be sizeable in
comparison to the symmetry-preserving one, $\mu_S^2$. In Fig.~\ref{fig:xSecs}
we show $\mu_X^2/\mu_S^2=0.1$ as a limit for
significant symmetry breaking.

\paragraph{One-loop contributions}

\begin{figure}[t]
    \begin{center}
	\includegraphics[width=0.45\textwidth]{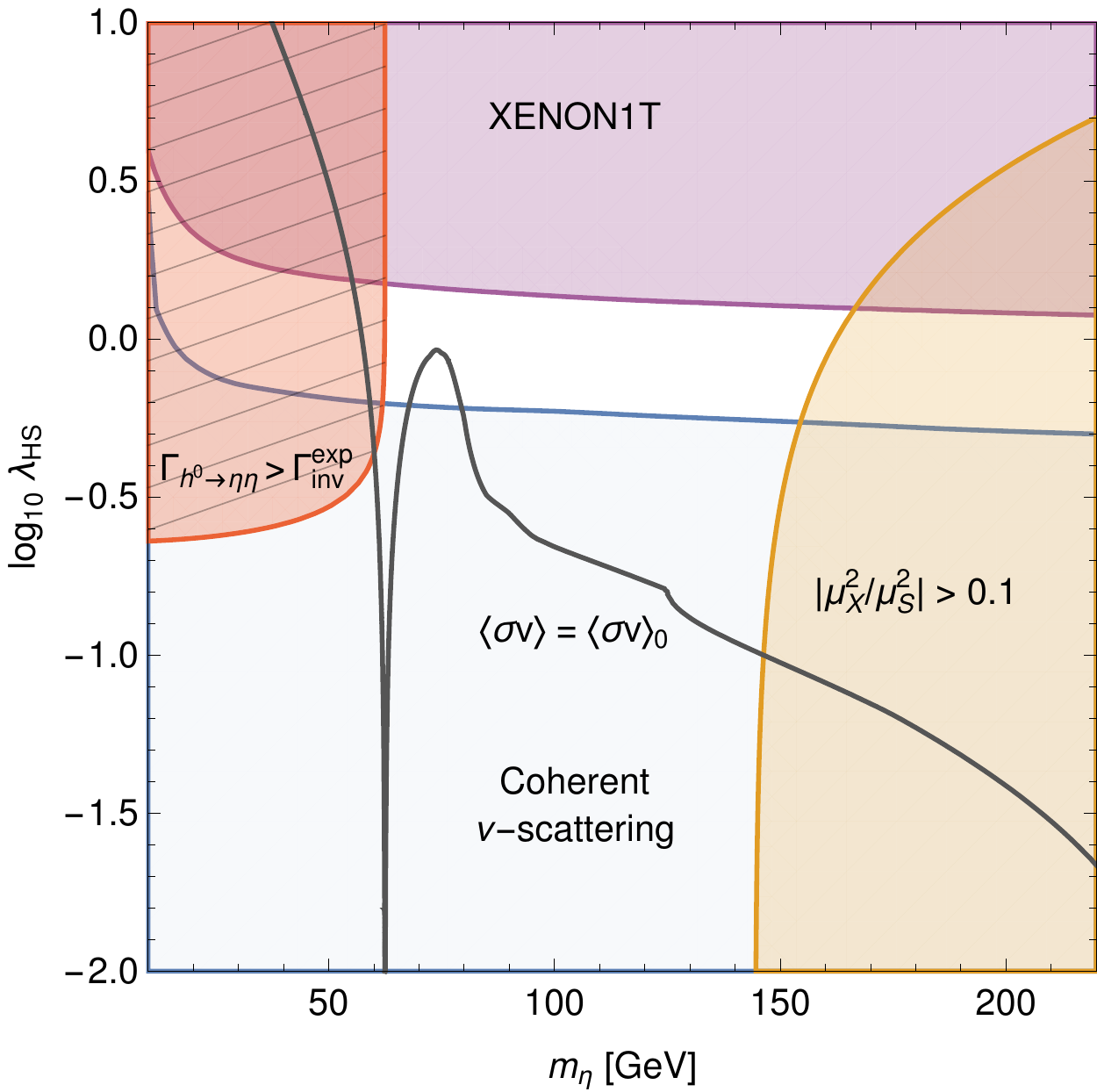}
    \end{center}
    \caption{The corresponding plot to Fig.~\ref{fig:xSecs}, with the dominant one-loop contributions taken into account, for the DM-nucleon cross section (see the text for details). We only plot here
    the ${\mH=500}$~GeV curves for clarity.
    }
    \label{fig:loop}
\end{figure}

The soft breaking of the global symmetry by $\mu_X^2$ does not affect the tree-level estimate of
the DM-nucleon scattering cross section, and the cross section remains momentum suppressed even in the presence of explicit breaking
of the symmetry. Therefore, one expects the quantum
corrections to induce contributions which scale as the DM mass, $m_\eta$, rather than the momentum transfer, $t$,
to the cross section manifesting the explicit breaking and the
pseudo-Goldstone nature of the DM candidate. This, indeed, turns out to be the case, and the one-loop contribution to the
DM-nucleon scattering has a piece independent of the momentum transfer which vanishes in the limit of $m_\eta\to0$.
This contribution dominates the DM-nucleon scattering by several orders of magnitude, and was estimated in Ref.~\cite{Gross:2017dan} to be
\begin{equation}
    \label{eq:1loopEst}
    \sigma_{\mathrm{SI}}^{1-\mathrm{loop}}\approx \frac{\sin^2\alpha}{64\pi^5}\frac{m_N^4f_N^2}{\mh^4v^2}\frac{\mH^2m_\eta^2}{w^6},
\end{equation}
for $m_\eta<\mH$.
The first exact one-loop computations~\cite{Azevedo:2018exj,Ishiwata:2018sdi} point to the same
ballpark region, indicating that Eq.~\eqref{eq:1loopEst} only slightly over-estimates the cross section. For the purpose of our
analysis, we use the conservative estimate of Eq.~\eqref{eq:1loopEst}, noting that an order-of-magnitude decrease in
the cross section does not alter the qualitative picture.

We show the corresponding plot to Fig.~\ref{fig:xSecs}, with the dominant one-loop contribution taken into account,
in Fig.~\ref{fig:loop}. We only show the ${\mH=500}$~GeV curves for simplicity, and refer to Fig.~\ref{fig:xSecs} for the change
due to increasing the heavy-scalar mass. The other parameter values are as in Fig.~\ref{fig:xSecs}.

\paragraph{Explicit breaking beyond mass terms}

In the presence of the quartic explicit-breaking coupling, $\lambda_X$, already the tree-level DM-nucleon scattering
cross section is sensitive to the breaking, and has a piece that does not vanish at zero momentum transfer: the second term in Eq.~\eqref{eq:lambdaeff}.

\begin{figure}[t]
    \begin{center}
	\includegraphics[width=0.45\textwidth]{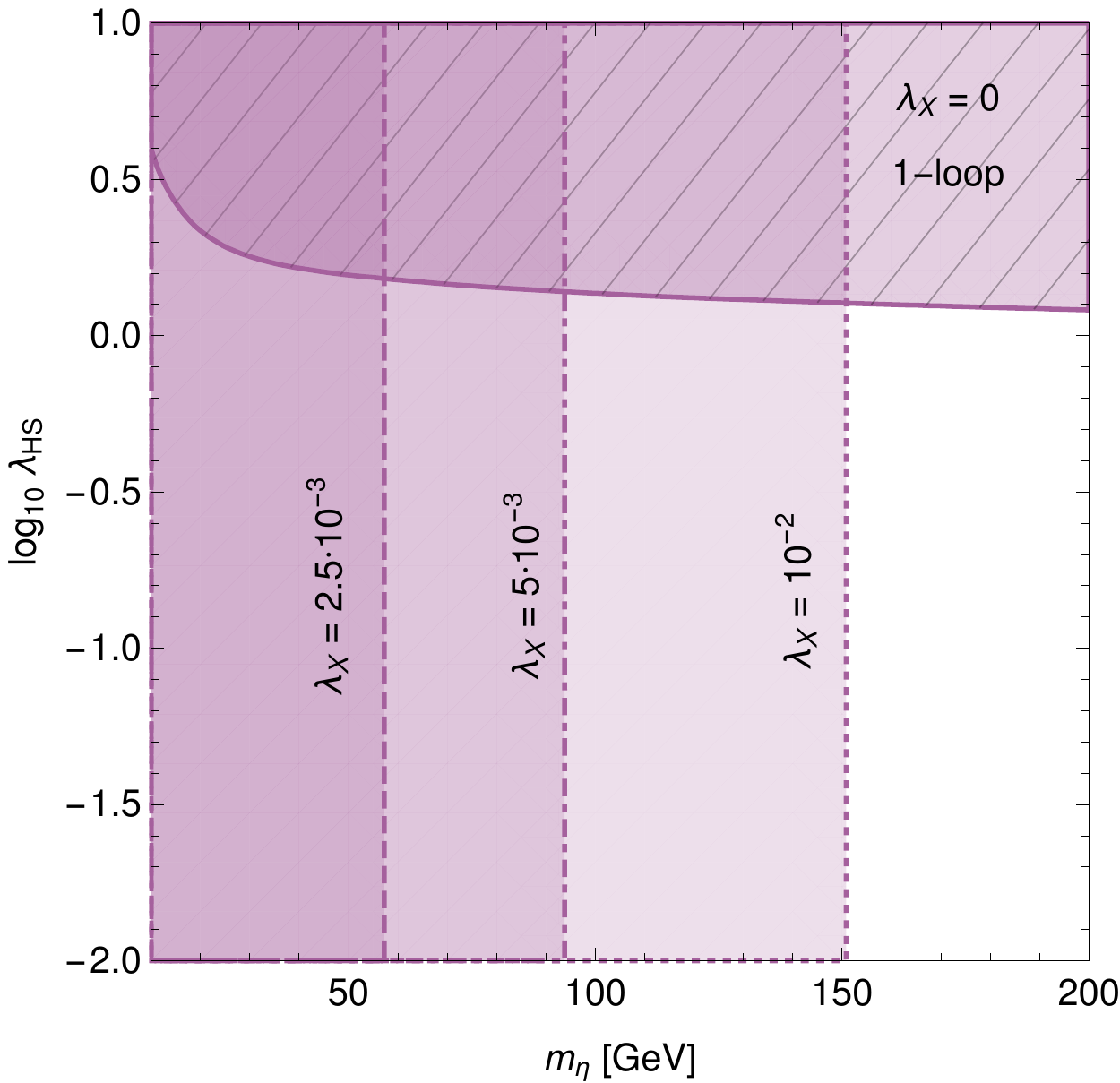}\quad
	\includegraphics[width=0.45\textwidth]{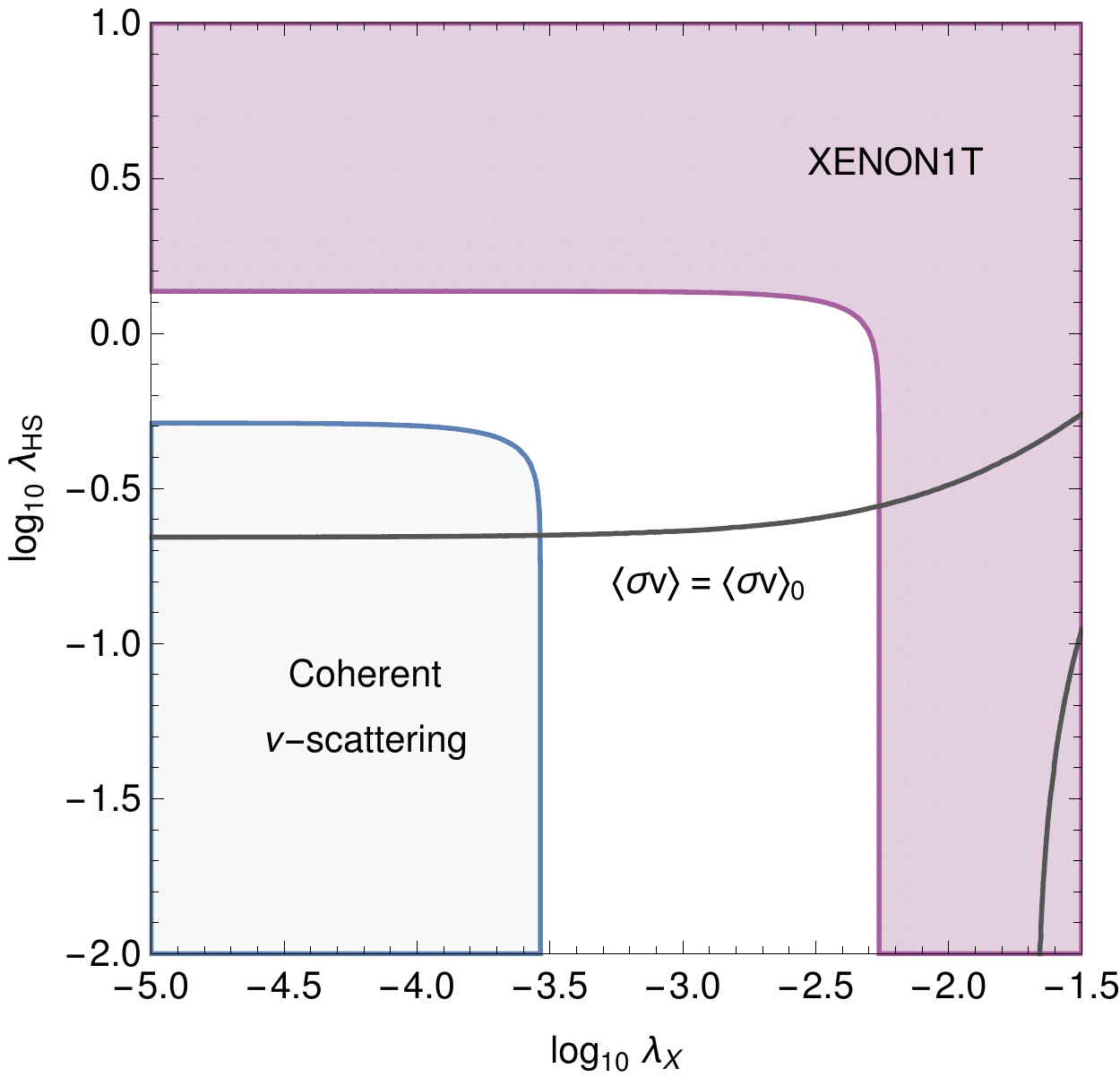}
    \end{center}
    \caption{{\bf Top panel:} the current limits from XENON1T on the spin-independent scattering cross section of
	DM off of nuclei for non-zero values of $\lambda_X$. {\bf Bottom panel:} the limits from XENON1T and the
	neutrino floor in the $(\lambda_{X},\lambda_{HS})$ plane for a fixed DM mass of $m_\eta=100$~GeV and $\alpha=0.3$.
	The gray curve shows where the observed DM abundance is obtained.}
    \label{fig:lambdaX}
\end{figure}

In the top panel of Fig.~\ref{fig:lambdaX}, we show how the XENON1T limit changes if the explicit-breaking parameter,
$\lambda_X$, is turned on; we plot the exclusion regions for fixed values of $\lambda_X=0, \; 2.5\cdot 10^{-3}, \; 5\cdot 10^{-3}, \;
10^{-2}$.
The bottom panel of Fig.~\ref{fig:lambdaX}, shows the XENON1T exclusion and the neutrino floor
along with the curve $\langle \sigma v_{\mathrm{rel}}\rangle=\langle \sigma v_{\mathrm{rel}}\rangle_0$ 
in the $(\lambda_X,\lambda_{HS})$ plane for fixed values of $m_\eta=100$~GeV
and $\alpha=0.3$. Changing $\alpha$ has little effect on the SI cross section, and thus varying $\alpha$ does not
noticeably change the plots.
These figures show the current direct-detection bounds can be evaded even if the
symmetry is only approximate or accidental; for example, for  ${m_{\eta}\sim150}$\,GeV non-negligible couplings of ${\lambda_X\lesssim 0.01}$ are still allowed.

%%INDIRECT DETECTION

\subsection{Indirect detection}
\begin{figure}[hbt]
  \includegraphics[width=0.45\textwidth]
  {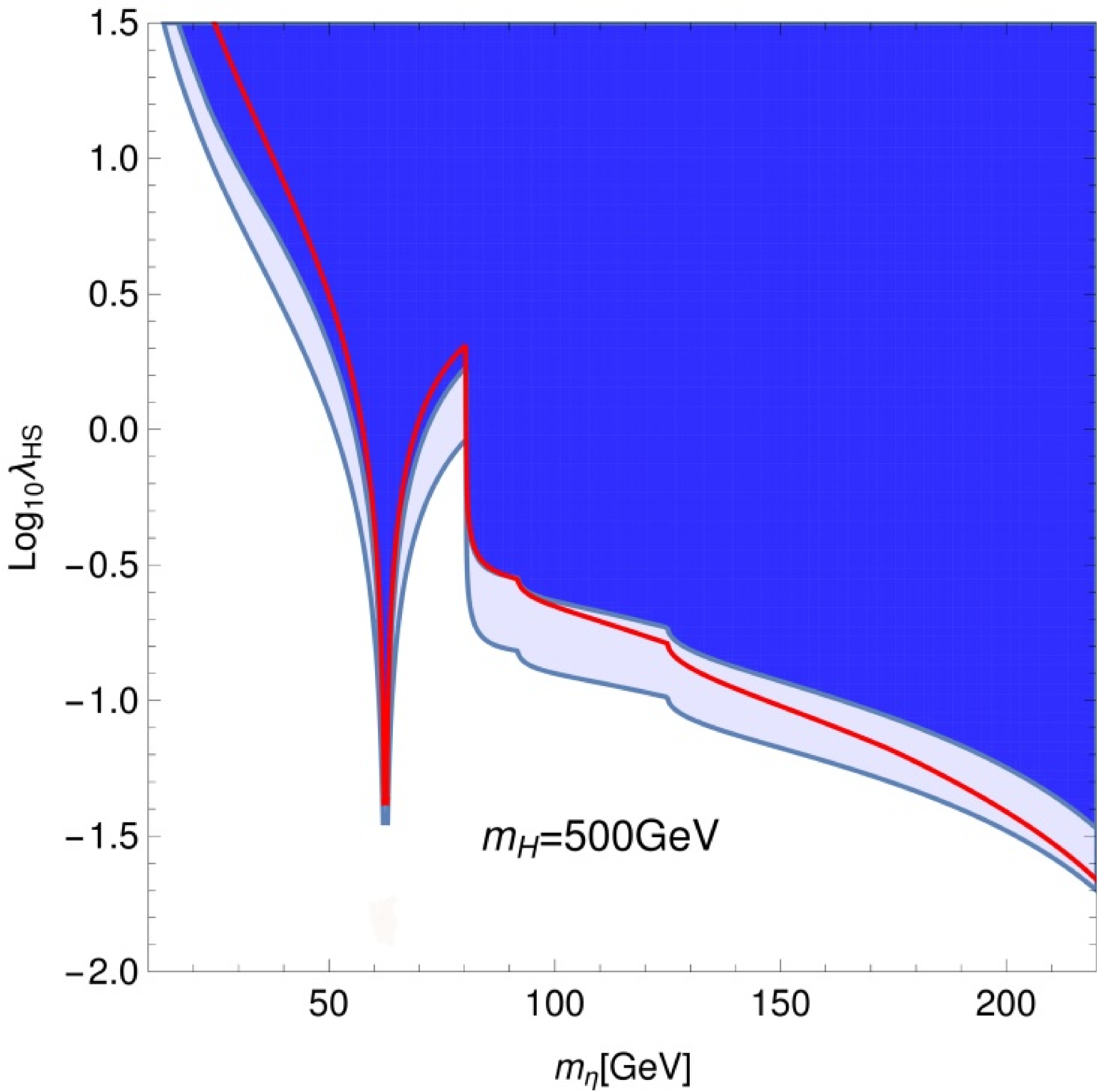}
  \includegraphics[width=0.45\textwidth]
  {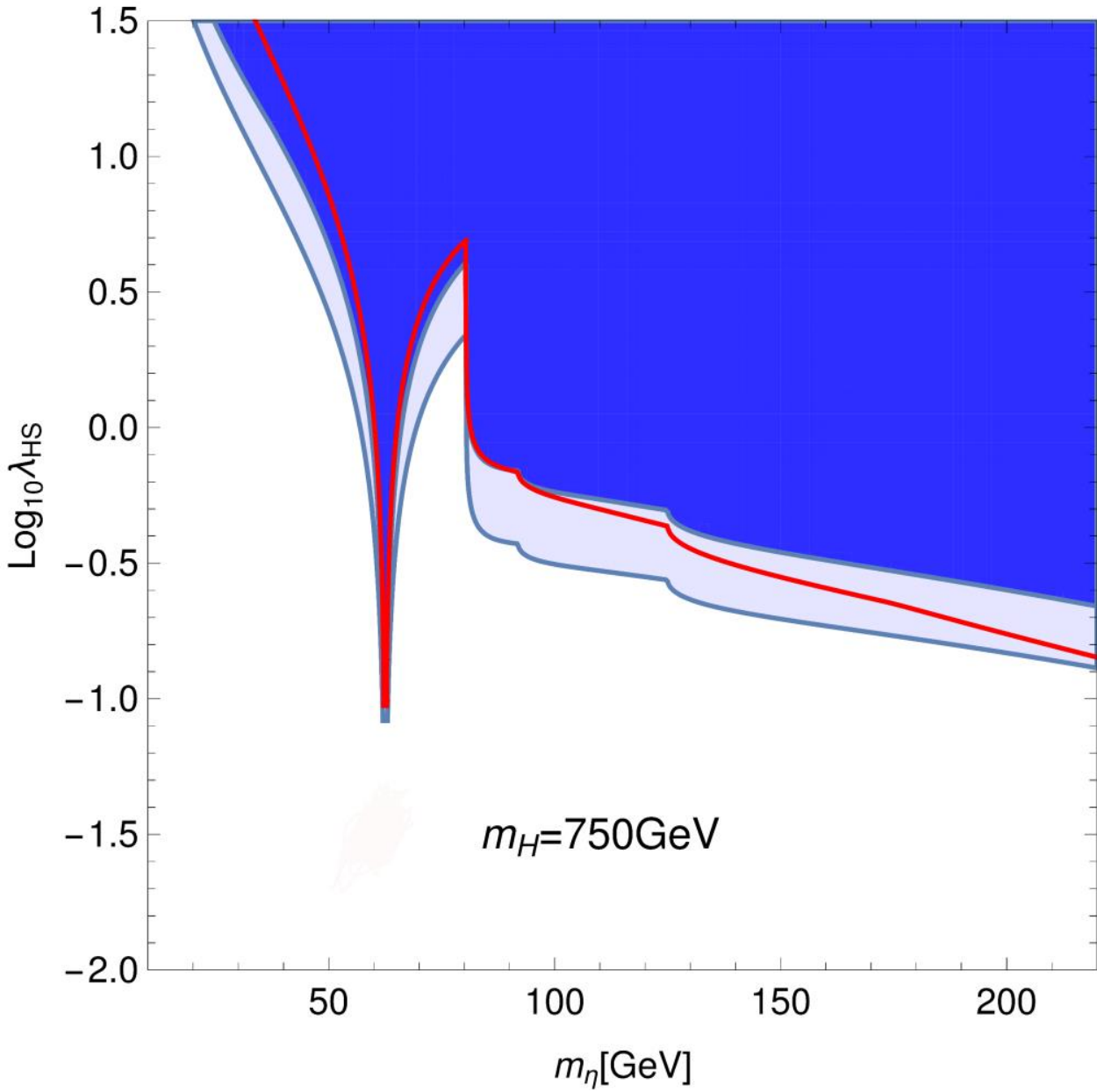}
  \caption{{\bf{Top panel:}} the direct-detection constraints for $\mH=500$\,GeV. {\bf{Bottom panel:}} same as the left panel, but for $\mH=750$\,GeV.}
\label{fig:indirect}
\end{figure}

The indirect-detection constraints arise from the annihilation of DM particles. The relevant cross section formulae are given in Appendix \ref{xsect}. To clarify the discussion, we show here just the annihilation cross section to the $\bar{b}b$-channel in the limit of $\lambda_X=0$:
\begin{equation}
v_{\mathrm{rel}}\cdot\sigma_{\eta\eta\to\bar{b}b}
=\frac{N_c \lambda_{HS}^2 \sqrt{s} \; m_b^2 (s-4m_b^2)^{3/2}}{4\pi(s-\mh^2)^2(s-\mH^2)^2}.
\end{equation}
Indeed, the cross section does not vanish in the non-relativistic limit of $s\rightarrow 4m_\eta^2$, provided that the process is kinematically allowed, $m_\eta>m_b$. Figure \ref{fig:indirect} shows the constraints from Fermi-LAT dwarf galaxy observations
applicable to the model, translated as an upper limit for the portal coupling,
$\lambda_{HS}$, as a function of the DM mass. Superimposed is the line where the
observed relic abundance is produced by freeze-out. The dark blue shaded region
is ruled out by the Fermi-LAT data~\cite{Fermi-LAT:2016uux}, and the light blue
shaded region is the expected exclusion region with 10 years of data, adapted
from Ref.~\cite{Charles:2016pgz}.

To produce the exclusion regions, we compare the
annihilation cross section $v\sigma(\eta\eta\rightarrow b\bar{b})$ with the
reported exclusion limit, for $m_\eta<m_W$. For $m_\eta>m_W$ the dominant
annihilation channel is $WW$, for which the constraints are not available in
the Fermi-LAT report. Based on model-independent analyses presented in
Ref.~\cite{Boddy:2018qur, Clark:2017fum}, we estimate the constraint for this channel to be
$v\sigma(\eta\eta\rightarrow WW) \approx 1.5 v\sigma(\eta\eta\rightarrow b\bar{b})$ in our mass range of interest. That is, we compare the model
prediction for the annihilation cross section in the $WW$ channel to the
reported upper limit for the $b\bar{b}$ cross section, scaled by a factor of $1.5$, for a given
value of the DM mass. As the DM mass increases further and the $ZZ$ and $h^0h^0$ channels become kinematically allowed, we also include the annihilation cross section into these channels with the same rescaling factor, i.e. we compare the sum of the cross sections $\eta\eta\rightarrow WW, ZZ, h^0h^0$, to the $\bar{b} b$ limit rescaled by a factor of $1.5$.

This procedure results in kinks in the
exclusion curve at $m_\eta=m_W, m_Z, \mh$. The kinks are non-physical, and would be
removed by including the contribution from the off-shell processes $\eta\eta \rightarrow WW^*, ZZ^* \rightarrow 4f$ etc., where $f$ is a SM fermion.
To give a more clear comparison between this limit and the thermal annihilation cross section, shown by the red line,
here we remove the full-Higgs-width correction also from the relic-abundance computation.
The smoothing around the kinematic thresholds is expected to be similar for both the exclusion region and the relic-abundance line. A more
detailed analysis of interpreting the Fermi-LAT constraints in terms of various
final states is beyond the scope of this work.

We conclude that with the treatment described above, the thermal relic cross section appears already excluded by the Fermi-LAT data for $m_{\eta}<m_W$, and will be probed up to $m_\eta \sim 250$\,GeV with future data. However, to make conclusive statements of the fate of the model in light of indirect detection constraints, a detailed analysis of the full set of annihilation final states is required.

\section{Conclusions}
\label{sec:checkout}

In this paper, we have considered a model framework which allows a paradigmatic
thermal relic WIMP to escape current direct-search limits and remain out of
reach also for future experiments. The effect arises due to momentum-dependent interactions between the DM and ordinary matter, and
appears naturally in scenarios where the DM candidate is a
pseudo-Goldstone boson of an approximate global symmetry~\cite{Hur:2011sv,Heikinheimo:2013fta,Azevedo:2018oxv,Balkin:2018tma}.

We first outlined the general formulation of this type of models and then, to
illustrate the effect more explicitly, considered how various observational constraints are implemented in the case  where the hidden sector
consists of a complex singlet scalar with an O(2) global symmetry. We
demonstrated that in this model one can obtain the correct DM relic density
and simultaneously remain out of the reach of current direct-search experiments.

In the tree-level analysis, the key feature is the
momentum dependence of the DM interactions. However, since the global symmetry is only approximate, the
effects of symmetry breaking must be addressed~\cite{Azevedo:2018exj}. A generic and concrete
origin of the slight violation of such symmetry are interactions with
high-energy degrees of freedom which are integrated out above the
energy scales of the effective theory for the DM and the SM.
We analysed how the symmetry-breaking operators
increase the direct-detection cross section. We found
that even when symmetry
breaking is present, the direct detection of such pseudo-Goldstone DM, requires
the future experiments to disentangle the signals of DM scattering from those
of the neutrino background.

As our main result, we established how this type of DM can be observed
by indirect detection. The annihilation cross section of the Goldstone boson DM
is not suppressed by the incoming three-momentum, and therefore
the constraints are similar to a vanilla WIMP candidate.
We conclude that the future Fermi-LAT observations will be able to discover or exclude this type of DM with masses up to few hundred GeV.

We showed the results for
a simple extension of the SM, where a Goldstone DM is consisted by a single scalar field. Our results can be
directly applied also to the case of a non-abelian symmetry
where the Goldstone DM arises as a degenerate multiplet of dark pions~\cite{Hur:2011sv,Hur:2007uz}: On one hand the direct detection will become more efficient as the event rate will increase proportionally to the DM degrees of freedom. On the other hand the indirect-detection constraint is independent on the DM multiplicity and the conclusions from the constraints
we have presented hold for any Goldstone DM within the class of models we considered in Sec.~\ref{sec:constraints}.

A further interesting extension to study would
be a scenario where the SM scalar sector is enlarged so that both DM and the Higgs
boson arise as pseudo-Goldstone particles~\cite{Alanne:2014kea,Alanne:2018xli}. In this type of models, the
interactions between the DM and ordinary matter would be 
further suppressed, and the origin of visible and dark matter would become more tightly tied together.

Finally, it would be interesting to study these models at finite temperature.
This would lead to determination of the consequences of such frameworks for
the electroweak phase transition and associated possibilities for electroweak
baryogenesis. A possible first-order phase transition in the hidden sector
would also lead to generation of gravitational wave signals~\cite{Schwaller:2015tja} possibly detectable
with future detectors.

\acknowledgments
This work has been financially supported by the Academy of
Finland project 310130. V K acknowledges the H2020-MSCA-RISE-2014 grant no. 645722 (NonMinimalHiggs) and the National Science Centre, Poland, the HARMONIA project under contract UMO-2015/18/M/ST2/00518 (2016-2019) and the discussion that followed the HARMONIA meeting. N K acknowledges financial support from the V\"ais\"al\"a foundation.

%%%%%%%%%%%%%%%%%%%%%%%%%%%%%%%%%%%%%%%%%%%%%
\appendix

\section{Cross sections}
\label{xsect}

Here we give the formulae for the computation of the annihilation cross section for the model considered in Sec. \ref{sec:model}. To make the equations more concise, it is useful to define the couplings
\begin{equation}
\begin{split}
\lambda_{\eta\eta h^0 h^0} =&\, (\lambda_{HS}+\lambda_X)c_\alpha^2+2\lambda_S s_\alpha^2, \\
\lambda_{\eta\eta H^0 H^0} =&\, (\lambda_{HS}+\lambda_X)s_\alpha^2+2\lambda_S c_\alpha^2,  \\
\lambda_{\eta\eta h^0} =&\, (\lambda_{HS}+\lambda_X) v c_\alpha-2\lambda_S w s_\alpha, \\
\lambda_{\eta\eta H^0} =&\, (\lambda_{HS}+\lambda_X) v s_\alpha+2\lambda_S w c_\alpha,\\
\lambda_{h^0h^0h^0}  =&\, 6 \lambda_H  v c_\alpha^3+3 (\lambda_{HS}-\lambda_X) v s_\alpha^2 c_\alpha \\
    & - 3(\lambda_{HS}-\lambda_X) w s_\alpha c_\alpha^2 - 6 \lambda_S w s_\alpha^3,\\
\lambda_{h^0h^0H^0} =&\,\frac{1}{4}\left[ (6\lambda_H+\lambda_{HS}-\lambda_X) v s_\alpha\right.\\
    &\quad +(6\lambda_S+\lambda_{HS}-\lambda_X) w c_\alpha\\
    &\quad +3(2\lambda_H-\lambda_{HS}+\lambda_X) v s_{3\alpha}\\
    &\left.\quad +3(\lambda_{HS}-2\lambda_{S}-\lambda_X) w c_{3\alpha}\right],\\
\lambda_{h^0H^0H^0} =&\,\frac{1}{4}\left[ (6\lambda_H+\lambda_{HS}-\lambda_X) v c_\alpha\right.\\
    &\quad-(6\lambda_S+\lambda_{HS}-\lambda_X) w s_\alpha\\
&\quad-3(2\lambda_H-\lambda_{HS}+\lambda_X) v c_{3\alpha}\\
    &\left.\quad+3(\lambda_{HS}-2\lambda_{S}-\lambda_X) w s_{3\alpha}\right],\\
\lambda_{H^0H^0H^0} =&\, 6\lambda_H v s_\alpha^3 +3 (\lambda_{HS}-\lambda_X) v s_\alpha c_\alpha^2 \\
    &+3 (\lambda_{HS}-\lambda_X) w s_\alpha^2 c_\alpha+6\lambda_{S} w c_\alpha^3,
\end{split}
\end{equation}
and
\begin{equation}\begin{split}
&\quad y_{h^0} = \frac{m_f}{v}c_\alpha, \quad y_{H^0} = \frac{m_f}{v}s_\alpha, \\
&g_{h^0Z} = \frac{v (g^2 + g'^2)}{2} c_\alpha, \quad g_{h^0W} = \frac{v g^2}{2} c_\alpha, \\
&g_{H^0Z} = \frac{v (g^2 + g'^2)}{2} s_\alpha, \quad g_{H^0W} = \frac{v g^2}{2} s_\alpha,
\end{split}\end{equation}
where we have used the short-hand notations $s_x\equiv \sin x$ and $c_x\equiv \cos x$. 
The annihilation cross section to fermion final states is
\begin{align}
\sigma_{\eta\eta\to\bar{f}f}
=&\frac{N_cm_f^2 \beta_f^{3}}{8\pi\beta_{\eta}(s-\mh^2)^2(s-\mH^2)^2}\label{eq:xSecff}\\
&\times\left[(\lambda_{HS}+\lambda_X)s-2\lambda_X\left(s_\alpha^2\mh^2+c_\alpha^2\mH^2\right)\right]^2,\nonumber
\end{align}
where
\begin{equation}
\beta_x\equiv\sqrt{1-\frac{4m_x^2}{s}}.
\end{equation}
The annihilation cross section to EW vector boson final states is
\begin{align}
    \sigma_{\eta\eta\to VV} &=\frac{\delta_V \beta_V \left(12m_V^4-4m_V^2 s+s^2\right)}{16\pi s \beta_{\eta}(s-\mh^2)^2(s-\mH^2)^2}  \\
	&\quad \times \left[(\lambda_{HS}+\lambda_X)s-2\lambda_X\left(s_\alpha^2\mh^2+c_\alpha^2\mH^2\right)\right]^2, \nonumber
\end{align}
where  $\delta_V=1$, $1/2$ for $W^\pm$ and $Z$ boson final states, respectively.

Finally, the cross sections for annihilation into
$h^0h^0$ final state is
\begin{align}
    & \sigma_{\eta\eta\to h^0h^0}=\frac{\beta_{h^0}}{32\pi s \beta_{\eta}}\left[\frac{16\lambda_{\eta\eta h^0}^4}{(s-2\mh^2)^2}
	\right. \nonumber\\
   &\left.+\left(\lambda_{\eta\eta h^0h^0}+\frac{\lambda_{h^0h^0h^0}\lambda_{\eta\eta h^0}}{s-\mh^2}
	+\frac{\lambda_{h^0h^0H^0}\lambda_{\eta\eta H^0}}{s-\mH^2}\right)^2\right.\label{eq:sigmahh}  \\
&\left.-\frac{8\lambda_{\eta\eta h^0}^2}{s-2\mh^2}
	\left(\lambda_{\eta\eta h^0h^0}+\frac{\lambda_{h^0h^0h^0}\lambda_{\eta\eta h^0}}{s-\mh^2}
	+\frac{\lambda_{h^0h^0H^0}\lambda_{\eta\eta H^0}}{s-\mH^2}\right)\right].\nonumber
\end{align}

In our scenario, the pseudo-Goldstone boson, $\eta$, is lighter than the scalar $H^0$, and we do not consider final states $h^0H^0$ and
$H^0H^0$. However, in a general set-up, the cross section for $H^0H^0$ final state can be obtained from
Eq.~\eqref{eq:sigmahh} by exchanging $h^0\leftrightarrow H^0$. The cross section for the mixed final
state $h^0H^0$ is lengthy and not particularly
illuminating, and we do not report that explicitly here.

\section{The non-linear representation treatment}
\label{sec:nonlinear}

In the non-linear representation, our starting point is the Lagrangian
\begin{equation}
{\cal L}=\frac{1}{2}{\rm{Tr}}(\partial_\mu S^\dagger \partial^\mu S)-V(H,S),
\label{eq:generaltheoryApp}
\end{equation}
where
\begin{equation}
H=\begin{pmatrix} \pi^+\\ \frac{1}{\sqrt{2}}(v+\phi+i\pi^0)\end{pmatrix},\quad
S=(w+\sigma)e^{ i \eta/w}.
\end{equation}
The scalar potential is given by
\begin{align}
\label{eq:scalarpotApp}
V(H,S)=&\,\mu_H^2H^\dagger H+\frac{1}{2}\mu_S^2 |S|^2
+\lambda_H(H^\dagger H)^2\nonumber\\
&+\frac{\lambda_{HS}}{2}(H^\dagger H)|S|^2
+\frac{\lambda_S}{4} |S|^4\\
&-\frac{1}{4}\mu_X^2(S^2+S^{*\,2}),
\nonumber
\end{align}
where we have for simplicity set the quartic symmetry-breaking coupling, $\lambda_X$, to zero.

The minimization conditions are as in Eq.~\eqref{eq:minconds} for ${\lambda_X=0}$:
\begin{align}
    \mu_H^2 &=-\lambda_H v^2-\frac{1}{2}\lambda_{HS}w^2,\nonumber \\
    \mu_S^2 &=-\lambda_S w^2-\frac{1}{2}\lambda_{HS}v^2+\mu_X^2,
    \label{eq:mincondsApp}
\end{align}
and we again 
define the mass eigenstates via a rotation
\begin{equation}
    \label{eq:rotationApp}
    \left(\begin{array}{c}h^0\\H^0\end{array}\right)=
    \left(\begin{array}{cc}\cos\alpha & -\sin\alpha\\ \sin\alpha & \cos\alpha\end{array}\right)
    \left(\begin{array}{c}\phi\\ \sigma\end{array}\right),
\end{equation}
with the mixing angle defined as
\be 
\tan(2\alpha)=\frac{-\lambda_{HS} vw}{\lambda_H v^2-\lambda_S w^2}.
\ee

The parameters $\lambda_H$, $\lambda_S, \lambda_{HS}$, and $\mu^2_X$ can then be rewritten in terms of the masses of the scalars and 
their mixing angle:
\begin{align}
\lambda_H 
=&
 \frac{1}{4 v^2}\left[(\mH^2+ \mh^2)-(\mH^2-\mh^2)\cos(2\alpha) \right],
\nonumber\\
\lambda_S 
=&
 \frac{1}{4 w^2}\left[(\mH^2+ \mh^2)+(\mH^2-\mh^2)\cos(2\alpha)\right],
 \nonumber\\
\lambda_{HS} 
=& 
\frac{\sin(2\alpha)}{2v w}(\mH^2-\mh^2),\label{eq:parsNL}\\
\mu^2_X
=&
\frac{1}{2}m_\eta^2.\nonumber
\end{align}

\begin{figure}[H]
\begin{center}
\begin{tikzpicture}[thick,scale=1.0]
\draw[dashed] (0,0) -- node[black,above,sloped,yshift=-0.0cm,xshift=-0.0cm] {$\eta$} node[black,below,sloped,yshift=-0.0cm,xshift=-0.3cm] {$p_1\ \rightarrow$} (1,-1);
\draw[dashed] (1,-1) -- node[black,above,sloped,yshift=0.0cm,xshift=0.0cm] {$\eta$}  node[black,below,sloped,yshift=0.0cm,xshift=0.3cm] {$\rightarrow\ k_1$}(2,0);
\draw[dashed] (1,-1) -- node[black,above,yshift=-0.3cm,xshift=-0.7cm] {$h^0, H^0$} (1,-2);
\draw (0,-3) -- node[black,above,sloped,yshift=0.0cm,xshift=0.0cm] {$f_{\mathrm{SM}}$} (1,-2);
\draw (1,-2) -- node[black,above,sloped,yshift=0.0cm,xshift=0.0cm] {$f_{\mathrm{SM}}$} (2,-3);
\end{tikzpicture}
\hspace{5mm}
\begin{tikzpicture}[thick,scale=1.0]
\draw[dashed] (0,0) -- node[black,above,sloped,yshift=-0.0cm,xshift=-0.0cm] {$\eta$} node[black,below,sloped,yshift=-0.0cm,xshift=-0.3cm] {$p_1\ \rightarrow$} (1,-1);
\draw[dashed] (0,-2) -- node[black,below,sloped,yshift=0.0cm,xshift=0.0cm] {$\eta$}  node[black,above,sloped,yshift=-0.0cm,xshift=-0.3cm] {$p_2\ \rightarrow$}  (1,-1);
\draw[dashed] (1,-1) -- node[black,above,sloped,yshift=0.0cm,xshift=0.0cm] {$h^0,H^0$} (2,-1);
\draw (2,-1) -- node[black,above,sloped,yshift=0.0cm,xshift=0.0cm] {$f_{\mathrm{SM}}$} (3,0);
\draw (2,-1) -- node[black,below,sloped,yshift=0.0cm,xshift=0.0cm] {$f_{\mathrm{SM}}$} (3,-2);
\end{tikzpicture}
\caption{Relevant diagrams for direct and indirect detection.}
\label{diagrams}
\end{center}
\end{figure}
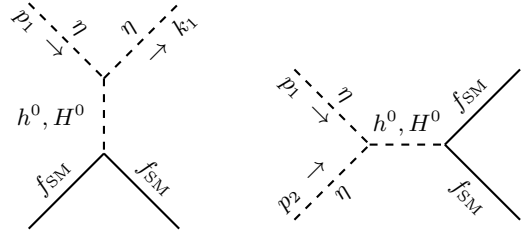

The relevant terms of the Lagrangian for direct and indirect detection and relic density calculations are
\begin{equation}
\begin{split}
V_{\mathrm{int}} 
\supset& 
-\frac{m_\eta^2}{w}\left[ \sin\alpha\; h^0 - \cos\alpha\; H^0 \right] \eta^2\\
 &+ 
\frac{1}{w} \left[ \sin\alpha \; h^0 - \cos\alpha \; H^0 \right]
(\partial_\mu \eta)^2,
\end{split}
\end{equation}
with the corresponding diagrams shown in figure \ref{diagrams}. For indirect detection we show here only the fermionic
final states for simplicity.

%%%%%%%%%%%%%%%%%%%%%%%%%%%%%%%%%%%%%%%%%%%%%
%%%%%%%%%%%%%%%%%%%%%%%%%%%%%%%%%%%%%%%%%

In the direct detection process, $\lambda_{\mathrm{ eff}}$ corresponding to that of Eqs~\eqref{eq:SIlinear}~and~\eqref{eq:lambdaeff}   has the form
\begin{align}  
\lambda_{ \mathrm{eff}}^{\mathrm{NL}}=&  
\frac{\sin(2\alpha)(\mh^2-\mH^2)}{vw (\mh^2-t)(\mH^2-t)}\left(-m_{\eta}^2-(-p_1\cdot k_1)\right)\nonumber\\
=&\frac{\sin(2\alpha)(\mh^2-\mH^2)}{vw (\mh^2-t)(\mH^2-t)}\left(-m_{\eta}^2-\frac{1}{2}(t-2m_{\eta}^2)\right)\nonumber\\
&=\ \frac{\sin(2\alpha)(\mh^2-\mH^2)}{vw (\mh^2-t)(\mH^2-t)}\cdot \frac{-t}{2}.
\end{align}
Writing $\lambda_{\mathrm{eff}}^{\mathrm{NL}}$ in terms of $\lambda_{HS}$ using Eq.~\eqref{eq:parsNL}, we obtain the same expression 
(up to an overall sign) as in the linear case, Eq.~\eqref{eq:lambdaeff}.

In the indirect detection, on the other hand, the corresponding effective coupling is given by
\begin{align}  
\lambda_{ \mathrm{eff, ID}}^{\mathrm{NL}}=&  
\frac{\sin(2\alpha)(\mh^2-\mH^2)}{vw (\mh^2-s)(\mH^2-s)}\left(-m_{\eta}^2-(p_1\cdot p_2)\right)\nonumber\\
=&\frac{\sin(2\alpha)(\mh^2-\mH^2)}{vw (\mh^2-s)(\mH^2-s)}\left(-m_{\eta}^2-\frac{1}{2}(s-2m_{\eta}^2)\right)\nonumber\\
&=\ \frac{\sin(2\alpha)(\mh^2-\mH^2)}{vw (\mh^2-s)(\mH^2-s)}\cdot \frac{-s}{2}.
\end{align}
Again, writing $\lambda_{\mathrm{eff, ID}}^{\mathrm{NL}}$ in terms of $\lambda_{HS}$ using Eq.~\eqref{eq:parsNL}, we recover the
same cross sections as in the linear case in the limit $\lambda_X=0$; see e.g. the cross section to fermionic final states, 
Eq.~\eqref{eq:xSecff}.

%%%%%%%%%%%%%%%%%%%%%%%%%%%%%%%%%%%%5
\bibliography{refs.bib}
%%%%%%%%%%%%%%%%%%%%%%%%%%%%%%%%%%%%5

\end{document}